\documentclass[twocolumn,amsmath,amssymb]{revtex4}
\usepackage{graphicx}% Include figure files
\usepackage{dcolumn}% Align table columns on decimal point
\usepackage{bm}% bold math
\begin{document}

\title{Disappearence of the Aharonov-Bohm Effect for Interacting Electrons in a ZnO
Quantum Ring}
\author{Tapash Chakraborty $^1$,\footnote{Tapash.Chakraborty@umanitoba.ca}
Aram Manaselyan$^2$, and Manuk Barseghyan$^2$}
\affiliation{$^1$ Department of Physics and Astronomy, University of
Manitoba, Winnipeg, Canada R3T 2N2}
\affiliation{$^2$ Department of Solid State Physics, Yerevan State
University, Yerevan, Armenia e-mails: amanasel@ysu.am, mbarsegh@ysu.am}
\date{\today}
\begin{abstract}
The electronic states and optical transitions of a ZnO quantum ring containing few interacting electrons 
in an applied magnetic field are found to be very different from those in a conventional semiconductor system, 
such as a GaAs ring. The strong Zeeman and Coulomb interaction of the ZnO system, exert a profound influence on 
the electron states and on the optical properties of the ring. In particular, our results indicate that 
the Aharonov-Bohm (AB) effect in a ZnO quantum ring strongly depends on the electron number. In fact, for two   
electrons in the ZnO ring, the AB oscillations become aperiodic, while for three electrons 
(interacting) the AB oscillations completely disappear. Therefore, unlike in conventional quantum ring
topology, here the AB effect (and the resulting persistent current) can be controlled by varying the electron number.

\end{abstract}
%\pacs{74.78.Na 73.63.Nm 74.45.+c 74.78.Fk}

\maketitle

In a quantum ring structure of nanoscale dimension, the confined electrons exhibit a topological quantum 
coherence, the celebrated Aharonov-Bohm (AB) effect \cite{AB_effect}. The characteristics of the energy 
spectrum (non-interacting) for a ring-shaped geometry, pierced by a magnetic flux $\Phi$, correspond to 
a periodically shifted parabola with period of one flux quantum, $\Phi^{}_0=h/e$ \cite{hund}. All physical 
properties of this system, most notably, the persistent current (magnetization) \cite{imry} and optical 
transitions \cite{tapash_optics}, have this periodicity. Experimental observations of the AB effect were 
reported in metal rings \cite{metal} and in semiconductor rings \cite{semicond}. Persistent currents were 
also measured in metal \cite{metal_2} and semiconductor \cite{semicond_2} rings. The role of electron-electron 
interactions on the AB effect was explored systematically via the exact diagonalization scheme for few 
interacting electrons in a quantum ring \cite{tapash_few,ring_review_theory}. Interactions were found to introduce 
fractional periodicity of the AB oscillations \cite{tapash_frac}. Major advances in fabrication of semiconductor 
nanostructures have resulted in creation of nanoscale quantum rings in e.g., GaAs and InAs systems containing 
only a few electrons \cite{ensslin_review,ring_review_expt}. In those experiments, the AB effect manifests
itself in optical transitions \cite{tapash_frac,lorke}, and magneto-conductance \cite{haug}. The electron 
energy spectrum in a ring geometry has also been measured \cite{ensslin}. Those experiments have confirmed 
the theoretical predictions about the influence of electron-electron interactions on the persistent current, 
that was previously predicted \cite{ensslin_review,ring_review_expt,ring_review_theory}. The AB effect 
has also been studied in Dirac materials, such as graphene \cite{abergeletal}, both theoretically 
\cite{tapash_graphene_ring} and experimentally \cite{graphene_ring}. One major advantage of all these 
nanoscale quantum rings is that here the ring size and the number of electrons in it can be externally 
controlled \cite{ensslin_review,ring_review_expt}.

In all these years, for investigations of nanoscale quantum structures, such as the  quantum dots (QDs) 
(or, the {\it artificial atoms}) \cite{Qdots,heitmann} and quantum rings (QRs), the materials of choice
had been primarily the conventional semiconductors, viz. the GaAs or InAs heterojunctions, where the 
high-mobility two-dimensional electron gas (2DEG) was quantum confined. In recent years, very exciting 
developments have taken place with the creation of high-mobility 2DEG in heterostructures involving 
insulating complex oxides. Unlike in traditional semiconductors, electrons in  these systems are strongly 
correlated \cite{mannhart}. These should then exhibit effects ranging from strong electron correlations, 
magnetism, interface superconductivity, tunable metal-insulator transitions, among others, and of course, 
the exciting possibility of all-oxide electronic devices. Many surprising results were found in the 
fractional quantum Hall states \cite{fqhe} discovered in the MgZnO/ZnO heterojunction \cite{falson,luo}. 
Preparation of various nanostructures, such as nanorings, nanobelts, etc. have been reported in ZnO 
\cite{zno_nano}. Here we report on the AB effect in a ZnO quantum rings and compare that in a conventional 
semiconductor QR, namely in GaAs. Quite remarkably, we found that while in the non-interacting case the 
AB effect remains unaltered for both systems, the combination of strong Zeeman interaction and the strong 
Coulomb interaction, two signature effects of the ZnO 2DEG, make the AB effect disappear in ZnO QR for 
electron number larger than one.

We consider here a two-dimensional QR with inner radius $R^{}_1$ and outer radius $R^{}_2$ having cylindrical 
symmetry, containing few electrons, in a magnetic field applied in the growth direction. The Hamiltonian 
of our system then is
\begin{equation}
{\cal H}=\sum_i^{N^{}_e}{\cal H}_\mathrm{SP}^i+\frac12\sum_{i\neq
j}^{N^{}_e}V^{}_{ij}, \label{Ham2D}
\end{equation}
where $N^{}_e$ is the number of electrons in the QR, $V^{}_{ij}=e^2/\epsilon\left|\mathbf{r}^{}_i-\mathbf{r}^{}_j
\right|$ is the Coulomb interaction term, with dielectric constant of the ring material $\epsilon$, and ${\cal 
H}^{}_\mathrm{SP}$ is the single-particle Hamiltonian in the presence of an external perpendicular magnetic field.

\begin{equation}\label{Hsp}
{\cal H}^{}_\mathrm{SP}=\frac{1}{2m}\left(\textbf{p}-\frac{e}{c}\textbf{A}\right)^2
+V^{}_{conf}(r)+\frac{1}{2}g\mu^{}_BB\sigma^{}_z,
\end{equation}
where $\textbf{A}=B/2(-y,x,0)$ is the vector potential of the magnetic field, $m$ is the electron effective mass. 
We chose the confinement potential of the QR with infinitely high borders: $V^{}_{\rm conf}(r)=0$, if $R^{}_1\leq 
r\leq R^{}_2$ and infinity outside of the QR. This choice of the confinement potential is suitable for ZnO/MgZnO 
heterostructures due to the large values of the conduction band offset and the electron effective mass \cite{Handbook}. 
The last term of (\ref{Hsp}) is the Zeeman interaction.

We take as basis states the eigenfunctions of $H^{}_\mathrm{SP}$ for $B=0$. The eigenfunctions of this Hamiltonian 
then have the form \cite{AregPhysE}
\begin{eqnarray}
\phi^{}_{nl}(r,\theta)&=&\frac{C}{\sqrt{2\pi}}e^{\mathrm{i}l\theta}\left(J^{}_l(\gamma^{}_{nl}r)-\frac{J^{}_l
(\gamma^{}_{nl}R^{}_1)}{Y^{}_l(\gamma^{}_{nl}R^{}_1)}Y^{}_l(\gamma^{}_{nl}r)\right),
\label{circularBasis}
\end{eqnarray}
where $J^{}_l(r)$ and $Y^{}_l(r)$ are Bessel functions of the first and second kind respectively, $\gamma^{}_{nl}=
2mE^{}_{nl}/\hbar^2$, where $E^{}_{nl}$ are the eigenstates defined from the boundary conditions, the constant $C$ 
is determined from the normalization integral, and $n$ and $l$ are the radial and angular quantum numbers respectively. 
In order to evaluate the energy spectrum of the many-electron system, we need to digonalize the matrix of the
Hamiltonian (\ref{Ham2D}) in a basis of the slater determinants constructed from the single-electron wave functions 
\cite{ring_review_theory,AregPhysE}.

We have also considered here the intraband optical transitions in the conduction band. According to the Fermi golden 
rule the intensity of absorption in the dipole approximation is proportional to the square of the matrix element 
\cite{tapash_optics,pekka}
\begin{equation}\label{Dipole}
I=\langle f|\sum_{i=1}^N r^{}_i e^{\pm i \theta^{}_i}|i\rangle
\end{equation}
when the transition goes from the initial $N$-particle state $|i\rangle$ to the final state $|f\rangle$. In this 
work we always consider $|i\rangle$ to be the $N$-particle ground state. To evaluate (\ref{Dipole}) we need to 
calculate the dipole matrix elements between the one electron states $|n,l\rangle$ and $|n',l'\rangle$.
\begin{equation}
M=\int_{R^{}_1}^{R^{}_2}\int_0^{2\pi}\phi^{}_{nl}(r,\theta)(re^{\pm
i\theta})\phi^{}_{n'l'}(r,\theta)rdrd\theta.
\end{equation}
After the angular integration we arrive at the optical transition selection rule for the total angular momentum 
$L^{}_f=L^{}_i\pm1$.

The numerical studies were carried out for the ZnO QR with parameters $m=0.24m^{}_0$, $g=4.3$, $\epsilon=8.5$ 
\cite{Handbook}. For the purpose of comparison we have also presented similar studies for the GaAs QR with parameters 
$m=0.067m^{}_0$, $g=-0.44$, $\epsilon=13.18$ respectively \cite{pekka}. We consider here the two QRs of same sizes 
with radii $R^{}_1=10$nm and $R^{}_2=40$nm.

\begin{figure}
\includegraphics[width=8cm]{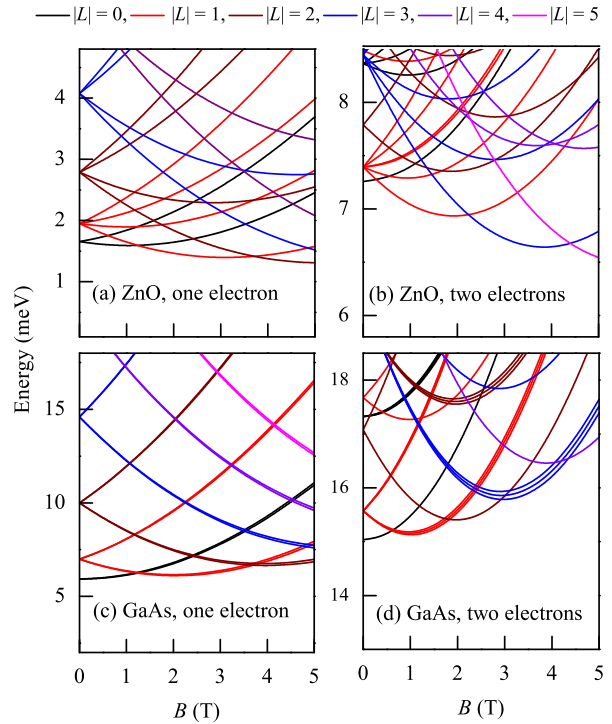}
\caption{\label{fig:EdepB1} The low-lying energy levels versus the magnetic field for (a) ZnO QR with 
one electron, (b) the ZnO QR with two electrons, (c) the GaAs QR with one electron and (d) the GaAs QR with 
two electrons. Different colors correspond to different values of the total angular momentum $L$.}
\end{figure}

The low-lying energy levels of the ZnO QR with one and two electrons are presented in Fig.~1 as a function of 
the magnetic field $B$. For comparison similar results are also presented for the GaAs QR  in Fig.~1(c) and (d). 
In all these figures different colors correspond to different values of the total angular momentum $L$ of the 
electrons. In the QR with only one electron in both systems, the ground state changes periodically with increasing 
magnetic field [Fig.~1(a) and (c)]. This is the direct signature of the AB effect in a QR. For the ZnO 
QR the energy eigenvalues are lower due to the larger value of the electron effective mass.  Additionally, the 
states with different spin are highly split  due to the larger value of the g-factor for ZnO. However, for the 
non-interacting electrons the AB effect survives in both systems.

For QRs with two interacting electrons, there are several substantial differences between the energy spectra of 
the ZnO and GaAs QRs. For instance, in the GaAs QR we see the usual and well observed AB oscillations due to 
level crossings between the singlet and triplet ground states, and for each crossing of the two-electron ground 
state the total angular momentum $L$ changes by unity. On the other hand, for the ZnO QR containing two electrons
[Fig.~1(b)], due to the combined effect of the strong Zeeman splitting and the strong Coulomb interaction, the 
singlet-triplet crossings disappear from the ground state. Interestingly, the periodic crossings happen only in
the excited states. For small values of the magnetic field the ground state is a singlet with $L=0$ and the total 
electron spin $S=0$. With an increase of the magnetic field the ground state changes to a triplet with $L=-1$ and 
$S=-1$. With further increase of the magnetic field all the observed crossings of the ground state correspond to 
triplet-triplet transitions between the states with odd number of total angular momentum ($|L|=1,3,5...$). These 
interesting and unexpected results will manifest themselves in optical transitions in the ZnO QR.

\begin{figure}
\includegraphics[width=8cm]{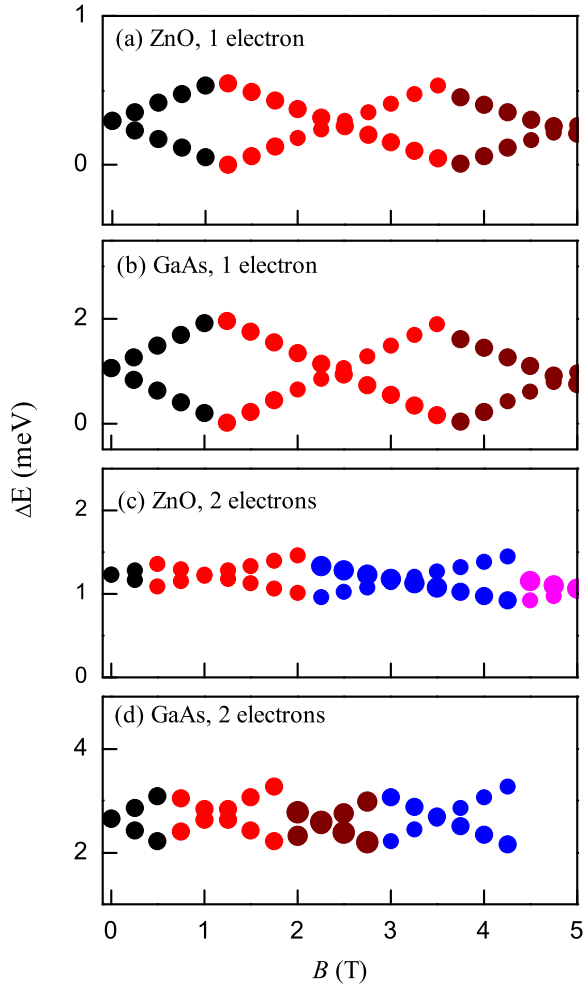}
\caption{Dipole allowed optical transition energies versus the magnetic field, for (a) the ZnO QR with one 
electron, (b) the GaAs QR with one electron, (c) the ZnO QR with two electrons and (d) the GaAs QR with two 
electrons. The size of the colored dots is proportional to the intensity of the calculated optical transitions.}
\end{figure}

The dipole allowed optical transition energies are presented in Fig.~2 as a function of the magnetic field for 
the ZnO ((a) and (c)) and the GaAs ((b) and (d)) QRs containing one and two electrons respectively. Different 
colors in Fig.~2 correspond to the value of the ground state angular momentum (see Fig.~1) of the optical
transition and the sizes of the points are proportional to the intensity of the optical transitions. For the QRs 
containing only one electron for both materials we can see the familiar picture: periodic optical AB oscillations. 
Comparing Fig.~2(a) and (b) we notice that although the strong Zeeman effect changes the one-electron energy 
spectra of the ZnO QR, it does not change the periodicity of the optical AB oscillations. In the case of the QRs 
with two electrons again we see considerable differences between the two systems. In the case of the two-electron 
GaAs QR, we see the periodic optical AB oscillations with the period that is half the flux quantum, which is an 
well-known result \cite{tapash_frac,haug}. In contrast, for the two-electron ZnO QR we notice an aperiodic behavior 
of optical AB oscillations. The first oscillation, which corresponds to the singlet-triplet transition from the 
state with $L=0$ to the state with $L=-1$ has a smaller period compared to the other oscillations which correspond 
to transitions between the triplet states with odd angular momentum. The period of these triplet-triplet oscillations 
is almost equal to the period of the single electron case. This unexpected effect is caused by the different properties 
of the energy spectra of the two-electron ZnO QR discussed above and can be explained by the combined effect of the 
strong Zeeman interaction and the strong electron-electron interaction in the ZnO.

\begin{figure}
\includegraphics[width=8cm]{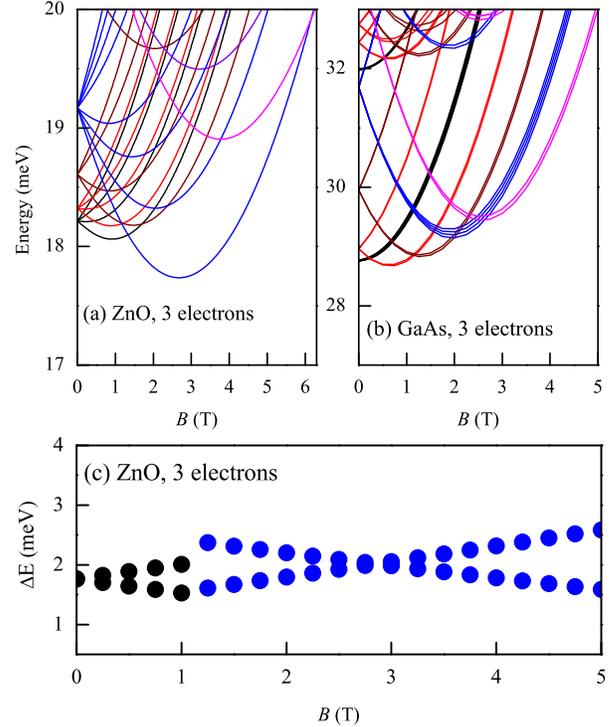}
\caption{Same as in Fig.~1 but for (a) the ZnO QR with three electrons and (b) the GaAs QR with three electrons. 
(c) Dipole allowed optical transition energies versus the magnetic field for the ZnO QR with three electrons.}
\end{figure}

In Fig.~3 (a) and (b) the low-lying energy levels for the ZnO and GaAs QRs containing three electrons are presented
as a function of the magnetic field $B$. For the three-electron GaAs QR we again note the periodic ground state
transitions and during each transition the ground state angular momentum changes by one. In contrast to that, for 
the three-electron ZnO QR only two ground state transitions are visible in that range of the magnetic field. At low 
magnetic fields the ground state has the angular momentum $L=0$. With the increase of the magnetic field at $B=1.3$T 
the ground state changes to $L=-3$. The next ground state transition appears at $B=6$T and the
angular momentum changes to $L=-5$. Therefore we can state that in the range of the magnetic field considered here
the Aharonov-Bohm effect disappears. The corresponding optical transition energies for the three-electron ZnO QR are 
shown in Fig.~3(c). That figure clearly illustrates the disappearence of the optical Aharonov-Bohm oscillations in
a ZnO quantum ring.

\begin{figure}
\includegraphics[width=8cm]{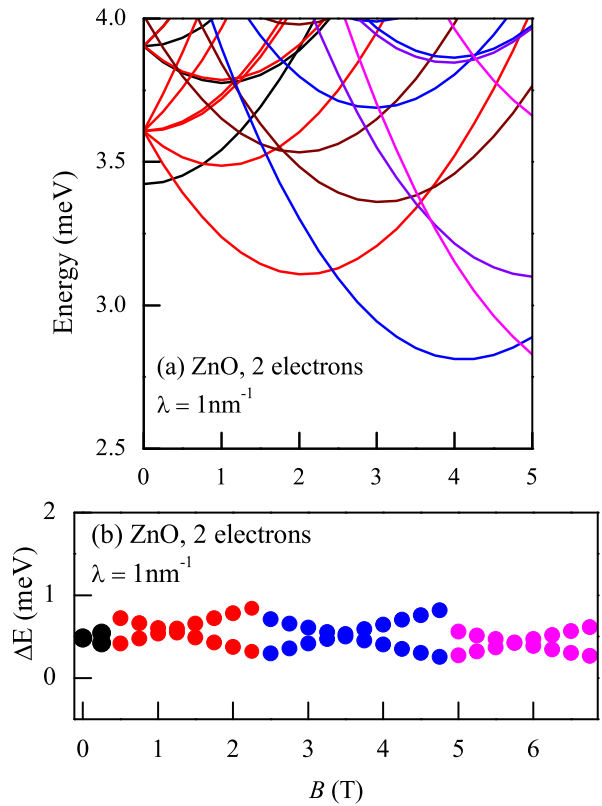}
\caption{Magnetic field dependence of the energy spectra (a) and the dipole allowed optical transition energies 
(b) for a two-electron ZnO QR with screened Coulomb interaction} \end{figure}

For a better understanding of the role of Coulomb interaction in the remarkable results for the ZnO QR shown above, 
we have considered also the case of two-electron ZnO QR with a screened Coulomb interaction. We have used the 
Yukawa-type screened Coulomb interaction potential \cite{majorana} $V^{}_{scr}=e^2e^{-\lambda |{\bf r}^{}_i-
{\bf r}^{}_j|}/\epsilon|{\bf r}^{}_i-{\bf r}^{}_j|$, where $\lambda$ is the screening parameter. The energy spectra 
and the optical transition energies are presented in Fig.~4 (a) and (b) respectively for $\lambda=1 {\rm nm}^{-1}$. 
The screened interaction shows that the first optical oscillation is observed for almost the same range of the magnetic 
field, as for the two-electron GaAs, but the periodicity is still absent. This clearly illustrates that interaction 
alone can not destroy the AB effect, but a strong Zeeman interaction is also important for that.

To summarize, we have studied the electronic states and optical transitions of a ZnO quantum ring containing a
few interacting electrons in an applied magnetic field via the exact diagonalization scheme. These results are
also compared with similar quantities for a conventional GaAs ring. We have found that the strong Zeeman interaction 
and the strong electron-electron Coulomb interaction, two major characteristics of the ZnO system, exert a profound
influence on the electron states and as a consequence, on the optical properties of the ring. In particular, 
we find that the AB effect is strongly electron number dependent. Our results indicate that in the case of two 
interacting electrons in the ZnO ring, the AB oscillations become aperiodic. For three electrons (interacting) we 
have found that the the AB oscillations actually disappear. These unusual properties of the ZnO QR are explained in 
terms of the energy level crossings that are very different from those of the conventional semiconductor QRs, such 
as for the GaAs. The AB effect (and thereby the persistent current) in a ZnO quantum ring can therefore be controlled 
by varying the electron number. 

The work has been supported by the Canada Research Chairs Program of the Government of Canada, the Armenian State 
Committee of Science (Project no. 15T- 1C331) and Armenian National Science and Education Fund (ANSEF Grant no. nano-4199).

\end{document}